\documentstyle[12pt]{article}
\newcommand{\z}{&\hspace*{-8pt}}

\begin{document}

\begin{center}
{\Large \bf Dispersive Approach to the Trace Anomaly.}\\[10mm]
{\bf \large N. Kawka}\\
{\it University of Vienna}\\
{\bf \large O. Teryaev {\rm and} O. Veretin}\\
{\it Bogoliubov Laboratory of Theoretical Physics,
            JINR, Dubna 141980, Russia} \\
\end{center}

\begin{abstract}
In the scalar $\phi^4$ field model the dispersive approach to the trace
anomaly is proposed. It is shown that it is impossible to get
dispersion representation for all formfactors so that preserve
both the translation and dilatation Ward identities.
Subtractions which preserve energy-momentum conservation
violate the classical trace Ward--Takahashi identity and
give rise to an anomalous contribution to the matrix element of
stress tensor $\theta_\mu^\nu$. This contribution coincides with
one-loop $\beta$-function in accordance with previous analyses based
on RG technique.
\end{abstract}
\vspace*{1cm}

   Some years ago in Ref. \cite{Dolgov} there was proposed
a derivation of the axial anomaly \cite{Adler} using dispersion
relations for formfactors. In this approach the anomaly appears
in a natural way as (finite) subtractions in dispersion
integrals. This formalism was developed in Ref. \cite{Achasov} for
different kinematical regimes. Another well known anomaly---that
of energy-momentum tensor trace \cite{Crewther}---had been
extensively studied in different field theories \cite{Duncan,Coleman,Brown}.
In this paper we apply the dispersion technique to the derivation
of the trace anomaly following the line of paper \cite{Dolgov}.

  In the case of the axial anomaly it was realized that there must be a
'pair' of symmetries (Ward identities at the quantum level) which
cannot be preserved under renormalization simultaneously and thus
give rise to anomaly. The trace anomaly, as it will be seen below, in fact
is a conflict between the translation and dilatation invariances.

  To derive the axial anomaly by dispersion method
one starts from the three-point Green's function
of one axial and two vector currents in a gauge theory with
a fermion field $\psi$
\begin{equation}
   T_{\alpha\mu\nu}(k_1,k_2) =
     \int {\rm d}x\, {\rm d}y\, {\rm e}^{i(xk_1+yk_2)}
     \langle J^5_\alpha(0) J_\mu(x) J_\nu(y)\rangle \,. \nonumber
\end{equation}
Conservation of the vector current
leads to the vector Ward identity
\begin{equation}\label{VWI}
   k_1^\mu T_{\alpha\mu\nu}=k_2^\nu T_{\alpha\mu\nu}=0\,,
\end{equation}
while (partial) conservation of the axial current is followed by
the axial Ward identity
\begin{equation}\label{AWI}
   (k_1+k_2)^\alpha T_{\alpha\mu\nu}=2m T_{\mu\nu}\,.
\end{equation}
Here $T_{\mu\nu}$ is the Green's function with the axial current replaced by
its divergence---chiral density $\bar\psi\gamma_5\psi$ and $m$ is the mass
of the fermion.

   It is well known that Ward identities (\ref{VWI}) and (\ref{AWI})
are incompatible at the one-loop level.
However, one can check that for the absorptive part
of $T$ both vector and axial identities hold true (no matter which
variable the discontinuity is taken with respect to). Using analytical
properties of amplitude $T$
dispersion relations for formfactors can be
written down. Then the anomaly manifests itself as additional
subtractions in dispersion integrals \cite{Dolgov,Achasov}.

One may think that such the scheme would work in the case of the
trace anomaly as well. Below we will show that it does (at the
one-loop level). As an example we considered
$\phi^4$ theory in four dimensions but generalizations of this
procedure on any scalar field theory is straightforward.

Let us start from the usual $\phi^4$-lagrangian of a real scalar
field $\phi$ with a mass $m$ and a coupling constant $\lambda$
\begin{equation}
  {\cal L} = \frac{1}{2}(\partial\phi)^2 - \frac{1}{2}m^2\phi^2
       - \frac{\lambda}{4!}\phi^4 \,.  \label{lagrangian}
\end{equation}

  At the classical level translations in the configuration space are
associated with the canonical stress tensor
$\theta_{\mu\nu}^{\rm can}$. However, in our analysis it is more convenient
to use the so-called 'improved' tensor \cite{Callan}
\begin{equation}
  \theta_{\mu\nu} = \frac12 (\partial_\mu\phi)(\partial_\nu\phi)
      + g_{\mu\nu} \frac12 m^2\phi^2
      + g_{\mu\nu} \frac{\lambda}{4!}\phi^4
      + \frac16 (g_{\mu\nu}\Box - \partial_\mu\partial_\nu)\phi^2\,,
    \label{tensor}
\end{equation}
which coincides with the canonical one $\theta_{\mu\nu}^{\rm can}$
apart from the last term. Both $\theta_{\mu\nu}$ and
$\theta_{\mu\nu}^{\rm can}$ generate space-time translations and have the
trace $m^2\phi^2+$total divergences (about an ambiguity
of energy-momentum tensor definition see e.g. review \cite{Coleman}).
Improved tensor (\ref{tensor}) has an advantage of having the trace
just $m^2\phi^2$ if equations of motion are applied
\begin{equation}
     \theta_\mu^\mu = m^2\phi^2 \,. \label{trace}
\end{equation}
Eq. (\ref{trace}) can be thought about as the 'classical trace identity'
while in the quantum theory the r.h.s. of (\ref{trace})
changes because of quantum loop corrections. One must consider
corresponding Ward identities instead of formula (\ref{trace}).
The translation and trace identities read \cite{Coleman}
\begin{eqnarray}
  q^\mu T_{\mu\nu}(p_1,\dots,p_n) \z=\z
   \sum_{i=1}^n (p_i+q)_\nu G^{(n)}(p_1,\dots,p_i+q,\dots,p_n)\,,
		   \label{translation WI}\\
  g^{\mu\nu} T_{\mu\nu}(q;p_1,\dots,p_n) \z=\z T(q;p_1,\dots,p_n)+
   \sum_{i=1}^n d\, G^{(n)}(p_1,\dots,p_i+q,\dots,p_n)\,.
		   \label{trace WI}
\end{eqnarray}
Here $G^{(n)}$ is the $n$-point Green's function having a canonical
dimension $d$ while $T_{\mu\nu}$ and $T$
are Green's functions of the operators $\theta_{\mu\nu}$ and
$\theta_\mu^\mu$ respectively, i.e.
\begin{eqnarray}
  G^{(n)}(p_1,\dots,p_n) \z=\z {\cal F}\langle
     T\phi(x_1)\dots\phi(x_n)\rangle\,,  \label{Gn}\\
  T_{\mu\nu}(q;p_1,\dots,p_n) \z=\z
    {\cal F}\langle T\theta_{\mu\nu}(0)\phi(x_1)\dots\phi(x_n)
            \rangle\,,  \label{Tmn}\\
  T(q;p_1,\dots,p_n) \z=\z
    {\cal F}\langle T\theta_\mu^\mu(0)\phi(x_1)\dots\phi(x_n)
            \rangle\,,  \label{T}
\end{eqnarray}
where symbol ${\cal F}$ stands for the Fourier transformation.
In the above formulae we reckon that all momenta are incoming.

  Consider first 2-point functions. Then $T_{\mu\nu}$ and $T$ are
operator insertions of $\theta_{\mu\nu}$ and $\theta_\mu^\mu$
into propagator $G^{(2)}$.
It was shown in Ref. \cite{Coleman} that there is no anomaly in this case
Namely, one can redefine tensor
$\theta_{\mu\nu}$ in such a way that both identities are fulfilled.
In another language it manifests the well known fact that there is no
mass renormalization to the first order in $\phi^4$ theory.

  Let us now turn to the case $n=4$ which we will deal with in the
rest of this paper. One-loop corrections to the matrix element
(\ref{Tmn}) are given by 9 diagrams.
But we are interested in those three only
that contribute to ${\rm Im}_{q^2}T_{\mu\nu}$. One of them
is shown in Fig.1 while two others are obtained by permutating
momenta $p_i$. Diagrams of other type corresponding
to $\phi^4$ terms in $\theta_{\mu\nu}$ are irrelevant when cut along $q^2$.

  The problem greatly simplifies if
we restrict ourselves to the following special kinematics
\footnote{This symmetrical kinematics is realized when the momenta
$\vec p_i$ in the c.m. frame $\vec q=0$ are directed
from the centre of the regular tetrahedron to its vertices.}
\begin{eqnarray}
  (p_i+p_j)^2 = p^2 < 0\, \qquad \mbox{for $i,j=1,...,4,\quad i\not=j$}.
       \label{kinematics}
\end{eqnarray}
In such a regime each of the three crossed diagrams gives
an equal contribution to formfactors depending only on two
variables $p^2$ and $q^2$. Thus one can consider only one graph keeping in
mind that the final result is to be multiplied by the factor $3$.

  Denoting the amplitude corresponding to the diagram of Fig.1 by
$\Delta_{\mu\nu}$ and introducing vectors $p=p_1+p_2$ and
$k=p_3+p_4$ we write the following decomposition
\begin{eqnarray}
   \Delta_{\mu\nu}(p,k) \z=\z  g_{\mu\nu} F_1
      + (p_\mu p_\nu + k_\mu k_\nu) F_2
      + (p_\mu k_\nu + k_\mu p_\nu) F_3\,,   \label{Delta}\\
   \Delta_\mu^\mu(p,k) \z=\z 2m^2 F_0\,,         \label{Delta1}
\end{eqnarray}
where formfactor $F_0$ corresponds to the triangle graph in Fig.1
with $\theta_{\mu\nu}$ replaced by trace operator $\theta^\mu_\mu$.
The same decomposition is valid for imaginary parts with replacements
$F_i\to A_i={\rm Im}F_i$.
The Feynman rules for operator vertexes are shown in Fig.2.
Calculation of the diagram of Fig.1 is straightforward and yields
\begin{eqnarray}
  A_1 \z=\z
   \frac{\lambda^2}{32\pi}\frac{1}{s+\beta}\Biggl\{
     \Bigl( \frac{s^2}{3} + \frac{s(2\beta+3)}{6}
    + \frac{\beta(\beta+4)}{8} \Bigr) L
   + \Bigl( s+\frac{\beta}{2} \Bigr) S \Biggr\}\,,  \label{A1}\\
  A_2 \z=\z \frac{\lambda^2}{32\pi}
   \frac{1}{4m^2(s+\beta)^2} \Biggl\{
    \Bigl( \frac{2s^2}{3} + \frac{s\beta}{3}
    - \frac{\beta(\beta+6)}{12} - \frac{\beta^2(\beta+4)}{8s} \Bigr) L
   +\Bigl( 2s-\frac{\beta^2}{2s}\Bigr) S \Biggr\}\,, \label{A2}\\
  A_3 \z=\z \frac{\lambda^2}{32\pi}
   \frac{1}{4m^2(s+\beta)^2} \Biggl\{
     \Bigl(-\frac{3s^2}{4} - \frac{s(5\beta+3)}{3}
      - \frac{\beta(5\beta+9)}{6} - \frac{\beta^2(\beta+4)}{8s}\Bigr) L
                            \nonumber\\
       \z+\z \Bigl(-4s-3\beta-\frac{\beta^2}{2s} \Bigr) S
               \Biggr\} \,, \label{A3}\\
  A_0 \z=\z \frac{\lambda^2}{32\pi} \frac{1}{2m^2} L\,,  \label{A0}
\end{eqnarray}
where we introduce the dimensionless variables
\begin{equation}
  s=\frac{q^2}{4m^2}\quad\mbox{and}\quad \beta=-\frac{p^2}{m^2}\,,
                        \label{vars}
\end{equation}
and
\begin{eqnarray}
  L \z=\z \frac{1}{\sqrt{s(s+\beta)}}
     \log\frac{s+\beta/2-\sqrt{(s-1)(s+\beta)}}
              {s+\beta/2+\sqrt{(s-1)(s+\beta)}}\,, \label{L}\\
  \qquad
  S \z=\z \sqrt{1-1/s}\,.       \label{S}
\end{eqnarray}

  Now let us consider Ward identities (\ref{translation WI}) and
(\ref{trace WI}) again. Substituting (\ref{Delta}),(\ref{Delta1})
into these equations we obtain the Ward
identities written in terms of formfactors
\begin{eqnarray}
  F_1(q^2,p^2)+\frac{q^2}{2} \Bigl( F_2(q^2,p^2)+F_3(q^2,p^2) \Bigr)
     \z=\z G^{(4)}(p^2)\,,     \label{F,translation}\\
  4F_1(q^2,p^2)+2p^2 F_2(q^2,p^2) + (q^2-2p^2) F_3(q^2,p^2) \z=\z
     2m^2 F_0(q^2,p^2)+4G^{(4)}(p^2)\,,
			 \label{F,trace}
\end{eqnarray}
where $G^{(4)}$ is the usual 4-point Green's function depending
only on $p^2$. Let us emphasize that there
are 'pure kinematical' contributions ($G^{(4)}$) in the r.h.s. of Eqs.
(\ref{F,translation}) and (\ref{F,trace}). Such terms are absent in
the axial--vector Ward identities. To get rid of the auxiliary
function $G^{(4)}$ one can introduce a new formfactor ${F'}_1=F_1-G^{(4)}$.
Then (\ref{F,translation}) and (\ref{F,trace}) take the form
\begin{eqnarray}
  {F'}_1(q^2,p^2)+\frac{q^2}{2} \Bigl( F_2(q^2,p^2)+F_3(q^2,p^2) \Bigr)
       \z=\z 0\,,    \label{F',translation}\\
4 {F'}_1(q^2,p^2)+2p^2 F_2(q^2,p^2)+ (q^2-2p^2) F_3(q^2,p^2)
       \z=\z 2m^2 F_0(q^2,p^2)\,.   \label{F',trace}
\end{eqnarray}
Furthermore, in our analysis we do not care much about $G^{(4)}$
for its discontinuity along $q^2$ axis vanishes.
For imaginary parts equations similar to
(\ref{F,translation}) and (\ref{F,trace}) exist
\begin{eqnarray}
  A_1(t,p^2)+\frac{t}{2} \Bigl( A_2(t,p^2)+A_3(t,p^2) \Bigr) \z=\z 0\,,
             \label{A,translation}\\
 4 A_1(t,p^2)+2p^2 A_2(t,p^2)+ (t-2p^2) A_3(t,p^2) \z=\z
     2m^2 A_0(t,p^2)\,,  \label{A,trace}
\end{eqnarray}
with $A_i(q^2)={\rm Im}_{q^2}F_i$.
It should be noted also that $A_1(t,p^2)\equiv{A'}_1(t,p^2)$.

  Another difference form axial
anomaly case is that ultraviolet divergences appear.
Namely, functions $F_1$ and $G^{(4)}$ are logarithmicaly
divergent while $F_2\,,F_3\,,F_0$ are finite.
One can convince oneself that functions $A_{2,3,0}(q^2)$ fall like
$1/q^2$ as $q^2\to\infty$. Thus we can postulate dispersion
representations for these formfactors without any subtractions
\begin{equation}
  F_{2,3,0}(q^2) = \frac{1}{\pi}\int\limits_{4m^2}^\infty
     \frac{A_{2,3,0}(t)}{t-q^2}\,{\rm d}t\,.  \label{DR}
\end{equation}
The situation is different for $A_1$. As one can expect
there is a logarithmic divergence and at least a single
subtraction is needed.
Let us make the subtraction at an euclidean point $q^2=\mu^2<0$
\begin{equation}
  F_1(q^2) =
    \frac{1}{\pi}(q^2-\mu^2) \int\limits_{4m^2}^\infty
    \frac{A_1(t)}{(t-\mu^2)(t-q^2)}\,{\rm d}t+c\,.  \label{F1ren}
\end{equation}
Note that generally $c$ is a function of $p^2$ and $\mu^2$.

   Integrating the Ward identities (\ref{A,translation})
and (\ref{A,trace}) we get for the translation identity
\begin{equation}
  F_1(\mu)+c+\frac{q^2}{2}(F_2+F_3)
    +\frac{1}{\pi}\int\limits_{4m^2}^\infty \left( \frac{A_1}{t-\mu^2}
       +\frac{1}{2}A_2+\frac{1}{2}A_3 \right) \,{\rm d}t = 0\,,
       \label{wi1}
\end{equation}
while the trace identity now reads
\begin{equation}
  4F_1(\mu)+4c+2p^2 F_2+(q^2-2p^2)F_3
    +\frac{1}{\pi}\int\limits_{4m^2}^\infty
    \left( \frac{4 A_1}{t-\mu^2}
       + A_3 \right) \,{\rm d}t = 2m^2 F_0\,.
       \label{wi2}
\end{equation}

From Eqs. (\ref{wi1}) and (\ref{wi2}) one can read that it is
impossible to satisfy both translation (\ref{F',translation})
and trace  (\ref{F',trace}) identities having the only auxiliary
parameter $c$. One is free though to choose $c$ so that
to preserve the former or the latter. It is more natural to have
the translation identity untouched, i.e. one should require that
Eq. (\ref{wi1}) takes form (\ref{F',translation}).
This fixes the subtraction $c(p^2)$:
\begin{equation}
  c(p^2) = -\frac{1}{\pi}
      \int\limits_{4m^2}^\infty \left( \frac{A_1}{t-\mu^2}
       -\frac{1}{2}A_2+\frac{1}{2}A_3 \right) \,{\rm d}t\,.
       \label{c}
\end{equation}

  Substituting (\ref{c}) into the trace identity (\ref{wi2}) we obtain
\begin{equation}
  F_1(\mu)+2p^2 F_2+(q^2-2p^2)F_3 = 2m^2 F_0 + \Delta\,,
      \label{traceanomaly}
\end{equation}
where $\Delta$ is, generally speaking, a function of $\mu^2$ and $p^2$ only.
\begin{eqnarray}
  \Delta(\mu^2,p^2) \z=\z -4c(\mu^2,p^2)
    -\frac{1}{\pi}\int\limits_{4m^2}^\infty \left(
   \frac{4 A_1}{t-\mu^2} + A_3 \right) \,{\rm d}t  \nonumber\\
   \z=\z \frac{1}{\pi}\int\limits_{4m^2}^\infty(2A_2+A_3)\,{\rm d}t\,.
                   \label{anomaly}
\end{eqnarray}
Let us emphasize that $\Delta$ in (\ref{anomaly}) is actually independent
on $\mu$, and by dimensional reasons (as it may depend only on ratio
$\mu^2/p^2$), on $p^2$.
All $\mu$-dependence is now absorbed in formfactor $F_1$ (its finite
renormalization actually changes it to $F'$).
Moreover, expressing $A_1$ in terms of $A_2$ and $A_3$ by use of the
translation invariance (\ref{A,translation})
and assuming the validity of (\ref{F',translation}), we may keep all the
derivation
ultraviolate finite on each stage.

Thus we see that the trace identity (\ref{traceanomaly}) develops the
extra term
$\Delta$ which is in fact the constant 'trace anomaly'.
Considering the chiral limit $m \to 0$, one may conclude, that because
of the constant value of the integral (\ref{anomaly}),
\begin{equation}
2A_2(t)+A_3(t) \to \pi \Delta \delta (t),
      \label{dilaton}
\end{equation}
manifesting the zero mass
"dilaton" singularity, in complete analogy with the axial anomaly case.
Trace anomaly is therefore  appearing as a purely infrared
phenomenon.

One may, in principle, consider the opposite case, when the dilatation
invariance is preserved, while the translation invariance is
anomalously broken. Expressing $A_1$ from (\ref{A,trace})
and assuming the validity of ({\ref{F',trace}) one immediately get:

\begin{eqnarray}
  {F'}_1(q^2,p^2)+\frac{q^2}{2} \Bigl( F_2(q^2,p^2)+F_3(q^2,p^2) \Bigr)
     \z=-\frac{1}{4}\Delta\,,     \label{translational anomaly}
\end{eqnarray}
the "translational anomaly" being described by the same expression
(\ref{anomaly})
as the trace one. It is leading to the similar massless singularity
(\ref{dilaton}) in the chiral limit, contrary to the case of
spontaneous breaking of translation invariance \cite{Jackiw}.

The integration in (\ref{anomaly}) can be performed analytically.
Evaluation yields that the result is really constant $\lambda^2/16\pi^2$.
Taking into account that the contribution to the Ward identity (\ref{trace WI})
is three times more because of the cross diagrams we obtain the anomalous
term in the r.h.s. of (\ref{trace WI})
\begin{equation}
  T^{\rm anom}= 3\Delta(\mu^2,p^2)=
    \frac{3\lambda^2}{16\pi^2}+O(\lambda^3)\,.   \label{Tanom}
\end{equation}

  Thus we have obtained the result which must be compared with
that given in Ref. \cite{Brown}. In terms of renormalized operators
the trace operator looks like \cite{Brown}
\begin{equation}
   \theta^\mu_\mu = m^2\phi^2+\beta(\lambda)\frac{1}{4!}\phi^4+\dots\,,
	\label{last}
\end{equation}
where dots stand for the terms which do not contribute to the matrix
element at hand and $\beta(\lambda)=3\lambda^2/16\pi^2+O(\lambda^3)$ is
the renormalization-group $\beta$-function. Taking matrix element
of (\ref{last}) one can see that $T^{\rm anom}$ is nothing but one-loop
$\beta$-function of the theory.
At the same time, the translation anomaly leads to:
\begin{equation}
 \partial_\mu  \theta^\mu_\nu =
 -\frac{1}{4}\beta(\lambda)\frac{1}{4!}\partial_\nu \phi^4.
         \label{last1}
\end{equation}
The equations (\ref{last},\ref{last1}) are representing the two different
choices of the anomaly for the pair of the symmetries and
may be compared with the
similar equations for the case of the another pair \cite{Jackiw2},
namely the general coordinate invariance and Weyl invariance
for the massless scalar
field interacting with the 2-dimensional gravity.
The r.h.s. of the analogies of (\ref{last},\ref{last1}) are $R/24\pi$ and
$\partial_\nu R/48\pi$, respectively, $R$ being the scalar curvature.
Note that the different magnitude of the ratio of the numerical
coefficients is the direct consequence of the different dimensionality
of the space-time.

When this investigation was finished, we have learned about
another efforts in this direction.

The generalization of our method to the more realistic case of QED
  and QCD appears to be rather lengthy, and the analysis of the
  Ward identities for the trace itself (assuming
  the translation invariance), successfully performed recently
  \cite{HoSt}, is more useful. However, we believe that the full analysis
  of the pair of anomalies, presented here, is also of some interest.

Also, the earlier calculations in the framework of the source theory
\cite{Milton} are in fact manifesting many properties of the
dispersive derivation and may be considered as its pioneering application
to the trace anomaly case.

At the same time,
the scalar fields are known to be of particular interest for the
quantum cosmology and general relativity,
where \cite{Deser} some analog of the
dispersive analysis (which seems to have a number of
counterparts in our approach)
was presented recently.

Authors are very thankful to A.P. Bakulev, A.V. Belitsky, A.T. Filippov,
J. Ho\v{r}ej\v{s}\'{\i}, S.V. Mikhailov, L. Sehgal and R.~Ruskov
for useful discussions,
and to R. Jackiw and K.~Milton for helpful correspondence.
This work is supported by RFFR grant N 96-02-17631.

                     %
                     %  F I G U R E S     1 and 2
                     %

\newpage
\vspace*{1cm}

                        % FIGURE 1

\begin{center}
\setlength{\unitlength}{1mm}
\begin{picture}(120,50)
%
% Triangle graph
%
\put(60,40){\circle*{2}}
\put(59.5,40){\line(0,1){4}}
\put(60.5,40){\line(0,1){4}}
%\put(60,42){\line(-1,1){3}}
%\put(60,42){\line(1,1){3}}

\put(56.5,42){$q$}
\put(62,40){$\theta_{\mu\nu}$}

\put(60,40){\line(-3,-5){20}}
\put(60,40){\line(3,-5){20}}
\put(40,6.5){\line(1,0){40}}

\put(40,6.5){\line(-1,0){6}}
\put(40,6.5){\line(0,-1){6}}
\put(80,6.5){\line(1,0){6}}
\put(80,6.5){\line(0,-1){6}}

\put(34,8){$p_1$}
\put(41,2){$p_2$}
\put(81,2){$p_3$}
\put(84,8){$p_4$}

%\capture{Fig.1}
\end{picture}
%\vskip -10mm
\centerline{
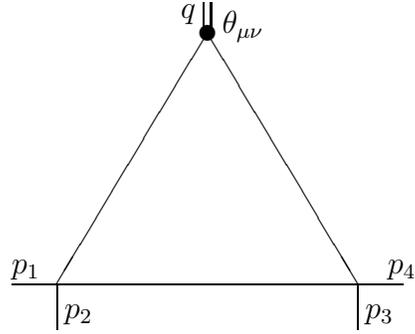
   {\bf Fig.1} One of the graphs contributing to the
   matrix element of $\theta_{\mu\nu}$.}
\end{center}
\vspace*{3cm}

                           % FIGURE2

\begin{center}
\setlength{\unitlength}{1mm}
\begin{picture}(120,45)
%
% Feynman rules for \theta_{\mu\nu} insertions
%
\put(30,40){\circle*{2}}
\put(30,40){\line(-1,0){15}}
\put(30,40){\line(1,0){15}}
\put(30,40){\vector(-1,0){10}}
\put(30,40){\vector(1,0){10}}
\put(18,37){$p_1$}
\put(42,37){$p_2$}
\put(28,43){$\theta_{\mu\nu}$}
\put(55,40){$i[g_{\mu\nu}(p_1 p_2 + m^2)-(p_{1\mu}p_{2\nu}+p_{1\nu}p_{2\mu})$}
\put(86,33){$+\frac13 (q_\mu q_\nu-q^2 g_{\mu\nu})]$}
%\put(85,30){$q = p_1 + p_2$}
%
\put(30,20){\circle*{2}}
\put(30,20){\line(-5,-3){15}}
\put(30,20){\line(5,3){15}}
\put(30,20){\line(-5,3){15}}
\put(30,20){\line(5,-3){15}}
%\put(30,20){\vector(-1,-1){10}}
%\put(30,20){\vector(1,1){10}}
%\put(30,20){\vector(-1,1){10}}
%\put(30,20){\vector(1,-1){10}}
\put(16,24){$p_1$}
\put(41,24){$p_2$}
\put(16,15){$p_3$}
\put(41,15){$p_4$}
\put(28,23){$\theta_{\mu\nu}$}
\put(55,20){$i\lambda g_{\mu\nu}$}
%\capture{Fig.2}
\end{picture}
\vskip -10mm
\centerline{
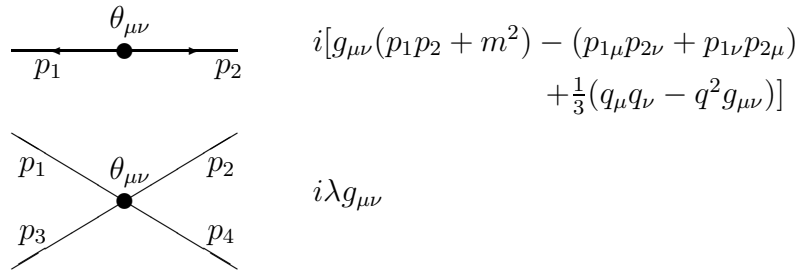
   {\bf Fig.2} The Feynman rules for operator $\theta_{\mu\nu}$.}
\end{center}


\begin{thebibliography}{99}
\bibitem{Dolgov}
A.D. Dolgov and V.I. Zakharov, Nucl.Phys. {\bf B27} (1971) 525.

\bibitem{Adler}   % axial anomaly introduced
S.L. Adler, Phys.Rev. {\bf 82} (1968) 664;\\
J.S. Bell and R. Jackiw, Nuovo Cim. {\bf A60} (1969) 47.

\bibitem{Achasov}
N.N. Achasov, Zh.Eksp.Teor.Fiz. {\bf 103} (1993) 11;\\
J. Ho\v{r}ej\v{s}\'{\i},  Phys.Rev. {\bf D32} (1985) 1029;\\
J. Ho\v{r}ej\v{s}\'{\i}  and O.V. Teryaev, Z. Phys. {\bf C65} (1995) 691;\\
O.V. Teryaev and O.L. Veretin, Yad.Fiz. {\bf 58} (1995) 2266.

\bibitem{Crewther} % dilatation anomaly introduced
R.L. Crewther, Phys.Rev.Lett. {\bf 28} (1972) 1421;\\
M.S. Chanowitz and J. Ellis, Phys.Lett. {\bf 40B} (1972) 397;\\
M.S. Chanowitz and J. Ellis, Phys.Rev. {\bf D7} (1973) 2490.

\bibitem{Duncan}  % trace anomaly in gauge theories
S.L. Adler, J.C. Collins and A. Duncan, Phys.Rev. {\bf D15} (1977) 1712;\\
J.C. Collins, A. Duncan and S.D. Joglekar, Phys.Rev. {\bf D16}
(1977) 438.

\bibitem{Coleman} % concerning WTI
S. Coleman and R. Jackiw, Ann.Phys. (N.Y.) {\bf 67} (1971) 552.

\bibitem{Callan} % the tensor
C. Callan, jr., S. Coleman and R. Jackiw, Ann.Phys. (N.Y.) {\bf 59} (1970) 42.

\bibitem{Brown} % composite operator technique
L.S. Brown, Ann.Phys. (N.Y.) {\bf 126} (1980) 135;\\
L.S. Brown and J.C. Collins, ibid. {\bf 130} (1980) 215.

\bibitem{Jackiw} E. D'Hoker and R. Jackiw, Phys. Rev. Lett {\bf 50}
(1983) 1719.

\bibitem{Jackiw2} R. Jackiw, Preprint MIT-CTP: \#2377 (November 1994).

\bibitem{HoSt}
J. Ho\v{r}ej\v{s}\'{\i}   and M. Stohr, Phys. Lett {\bf B379} (1996) 159;\\
J. Ho\v{r}ej\v{s}\'{\i}   and M. Schnabl, Charles University
preprint PRA-HEP 97/2 (hep-ph/9701397).

\bibitem{Milton} K. Milton, Phys.Rev. {\bf D7}
(1973) 1120; ibid., {\bf D15} (1977) 532, 2149.

\bibitem{Deser}
     S. Deser, Brandeis University preprint No. BRX TH - 399
     (hep-th/9609138), Appendix B.

\end{thebibliography}
\end{document}